\documentclass[iop,apj,tighten]{emulateapj}
\usepackage{amsmath,amstext}
\usepackage{apjfonts}
\usepackage[breaklinks,colorlinks,citecolor=blue,linkcolor=magenta]{hyperref} 
\usepackage[all]{hypcap} 
\usepackage{float}
\usepackage{footmisc}
\usepackage{multirow}
\usepackage{lineno}

\def \nustar {{\em NuSTAR}}

\def \swift {{\em Swift}}
\def \astrosat {{\em AstroSAT}}
\def \inte {{\em INTEGRAL}}
\def \swiftsource {{Swift J1658.2--4242}}

\shorttitle{\swiftsource\ Hard State}
\shortauthors{Xu et al.}

\begin{document}

\title{The Hard State of the Highly Absorbed High Inclination Black Hole Binary Candidate \swiftsource\ Observed by \nustar\ and \swift}

\author{Yanjun Xu\altaffilmark{1}}
\author{Fiona A. Harrison\altaffilmark{1}}
\author{Jamie A. Kennea\altaffilmark{2}}
\author{Dominic J. Walton\altaffilmark{3}}
\author{John A. Tomsick\altaffilmark{4}}
\author{Jon M. Miller\altaffilmark{5}}
\author{Didier Barret\altaffilmark{6,7}}
\author{Andrew C. Fabian\altaffilmark{3}}
\author{Karl Forster\altaffilmark{1}}
\author{Felix F\"urst\altaffilmark{8}}
\author{Poshak Gandhi\altaffilmark{9}}
\author{Javier A. Garc\'ia\altaffilmark{1,10}}

\altaffiltext{1}{Cahill Center for Astronomy and Astrophysics, California Institute of Technology, Pasadena, CA 91125, USA}
\altaffiltext{2}{Department of Astronomy and Astrophysics, Pennsylvania State University, 525 Davey Lab, University Park, PA 16802, USA}
\altaffiltext{3}{Institute of Astronomy, University of Cambridge, Madingley Road, Cambridge CB3 0HA, UK}
\altaffiltext{4}{Space Sciences Laboratory, 7 Gauss Way, University of California, Berkeley, CA 94720-7450, USA}
\altaffiltext{5}{Department of Astronomy, University of Michigan, 1085 South University Avenue, Ann Arbor, MI 48109, USA}
\altaffiltext{6}{Universite de Toulouse, UPS-OMP, IRAP, Toulouse, France}
\altaffiltext{7}{CNRS, IRAP, 9 Av. colonel Roche, BP 44346, F-31028 Toulouse cedex 4, France}
\altaffiltext{8}{European Space Astronomy Centre (ESA/ESAC), Operations Department, Villanueva de la Ca\~nada (Madrid), Spain}
\altaffiltext{9}{Department of Physics and Astronomy, University of Southampton, SO17 3RT, UK}
\altaffiltext{10}{Remeis Observatory \& ECAP, Universit\"at Erlangen-N\"urnberg, Sternwartstr.~7, 96049 Bamberg, Germany}

\begin{abstract}
We present a spectral and timing analysis of the newly reported Galactic X-ray transient \swiftsource\ observed by \nustar\ and \swift. The broad-band X-ray continuum is typical of a black hole binary in the bright hard state, with a photon index of $\Gamma=1.63\pm0.02$ and a low coronal temperature of $kT_{\rm e}=22\pm1$~keV, corresponding to a low spectral cutoff well constrained by \nustar. Spectral modeling of the relativistic disk reflection features, consisting of a broad Fe K$\alpha$ line and the Compton reflection hump, reveals that the black hole is rapidly spinning with the spin parameter of $a^{*}>0.96$, and the inner accretion disk is viewed at a high inclination angle of $i=64^{+2}_{-3}{^\circ}$ (statistical errors, 90\% confidence). The high inclination is independently confirmed by dips in the light curves, which can be explained by absorbing material located near the disk plane temporarily obscuring the central region. In addition, we detect an absorption line in the \nustar\ spectra centered at $7.03^{+0.04}_{-0.03}$~keV. If associated with ionized Fe K absorption lines, this provides evidence for the presence of outflowing material in the low/hard state of a black hole binary candidate. A timing analysis shows the presence of a type-C QPO in the power spectrum, with the frequency increasing from $\sim0.14$~Hz to $\sim0.21$~Hz during the single \nustar\ exposure. Our analysis reveals that \swiftsource\ displays characteristics typical for a black hole binary that is viewed at a high inclination angle, making it a good system for studying the accretion geometry in black hole binaries.

\end{abstract} 

\maketitle

\section{INTRODUCTION}
Most Galactic black hole binaries are discovered as X-ray transients that go into recurrent outbursts. During a typical outburst, a black hole binary transitions from the low/hard state (power-law component dominates the energy spectrum with a hard photon index) to the high/soft state (thermal disk component dominates) as the source flux increases \citep[see][for a review]{rem06}. The changes in the spectral shape are believed to be associated with the evolution of the accretion geometry. One hypothesis that has been widely explored is that the inner disk extends to the innermost stable circular orbit (ISCO) in the soft state, whereas is truncated at a larger radius in the hard state \citep[e.g.,][]{esin97,done07}. However, it has been suggested from observations that the inner disk extends down to the ISCO in the hard state of black hole binaries in several cases \citep[e.g.,][]{miller02,miller06}. Whether the accretion disk is truncated can be determined by measuring the degree of relativistic distortion of the Fe K$\alpha$ emission line, which comes from reflection of the central emission by the inner accretion disk \citep[e.g.,][]{fabian89}. Recently, \nustar\ observations of several black hole binaries in the bright hard state revealed very broad Fe K$\alpha$ lines \citep[][]{miller15, el_batal16, xu_maxi18}. As \nustar\ spectra are free from pile-up distortions even at high count rates typical of Galactic X-ray binaries, these results clearly challenge the disk truncation interpretation of the hard state.

Mass outflows are important phenomena in the study of accretion processes. Equatorial disk winds, identified by narrow absorption lines from highly ionized iron (mostly Fe {\small XXV} and Fe {\small XXVI}), have been observed in a number of black hole binaries in the soft state, and are more likely to be observed in high inclination systems \citep[e.g.,][]{neil09, king14, miller15_wind}. Detailed studies of several well-known black hole binaries indicate that the presence of accretion disk winds appears to be anti-correlated with relativistic jets \citep[e.g.,][]{neil09,miller12}. The disk wind is thought to be responsible for carrying away a considerable amount of kinetic energy when the jet switches off in the high/soft state. During typical low/hard states when the source Eddington ratio is relatively low, however, collimated radio jets are ubiquitous, and disk wind features are either absent or at least much weaker, thus making the detection difficult  \citep[e.g.,][]{ponti12,miller12}. 

\swiftsource\ is a newly discovered X-ray transient in the Galactic plane. The first reported detection was made by \swift/BAT \citep{swiftbat} on February 16, 2018. The source was detected by \inte\ during its observations of the Galactic center field on February 13, 2018, with a spectral hardness typical for black hole binaries in the hard state \citep{inte1, inte2}. Subsequent radio observations by ATCA imply that the source is a black hole binary at a distance of greater than 3~kpc \citep{russ18}. Low frequency quasi-periodic oscillations (QPOs) increasing in frequency were detected by \nustar\ \citep{xu18} and \astrosat\ \citep{beri18}. X-ray spectra from \swift/XRT reveal that the source is highly absorbed, with an absorption column density of $N_{\rm H}>10^{23}$~cm$^{-2}$ \citep{lien18}. In addition, several dips were observed from initial analysis of the \nustar\ data ~\citep{xu18}, similar to those detected in known dipping low-mass X-ray binaries (LMXBs) \citep[e.g.,][]{tom98, trigo09, kuu13}. Dips in the light curves of black hole and neutron star binaries are signatures of high inclination ($i>65{^\circ}$). They are thought to originate from obscuring material located at the thick outer region of the accretion disk in LMXBs \citep{white82,frank87}; while in HMXBs, dips can be explained by obscuration by the stellar wind from the donor star.

In this paper, we present a spectral and timing analysis of the newly discovered black hole binary candidate \swiftsource, using simultaneous \nustar\ and \swift\ observations that caught the source in the hard state. The paper is structured as follows: in Section \ref{sec:sec2}, we describe the observations and the data reduction details; we present the results of our spectral modeling and timing analysis in Section \ref{sec:sec3}, \ref{sec:sec4} and \ref{sec:sec5}, and discuss the results in Section \ref{sec:sec6}.

\section{OBSERVATIONS AND DATA REDUCTION}
\label{sec:sec2}
\swiftsource\ was observed by \nustar\ \citep{nustar} on February 16, 2018 starting at 23:26:09 UT for an exposure of 33.3~ks per module (OBSID: 90401307002). The observation was approved through DDT time, and was taken on the same day that the outburst was first reported. We reduced the \nustar\ data following standard procedure using NuSTARDAS pipeline v.1.6.0 and CALDB v20170817. The source spectra were extracted from a circular region with the radius of 150$\arcsec$ from the two \nustar\ focal plane modules (FPMA and FPMB). Corresponding background spectra were extracted using polygonal regions from source-free areas in the detectors. We also extracted spectra from mode 6 data following the procedures described in \cite{walton16}, which yielded an extra $\sim$3.6 ks of data when an aspect solution was not available from the on board star tracker CHU4. We grouped the \nustar\ spectra with a signal-to-noise ratio (S/N) of 20 per energy bin.

The outburst of \swiftsource\ was monitored by \swift/XRT \citep{swiftxrt}. The XRT observation (OBSID: 00810300002) overlapping with the \nustar\ observation used in this paper was taken in the Windowed Timing (WT) mode to avoid photon pile-up. The XRT observation was taken around the middle of the \nustar\ exposure, starting from 10:38:52 UT to 16:43:57 UT on February 17, 2018. We reduced the the \swift/XRT data using {\tt xrtpipeline} v.0.13.2 with CALDB v20171113. We extracted the source spectrum from a circular region with the radius of 70$\arcsec$, and the background was extracted from an annulus area with the inner and outer radii of 200$\arcsec$ and 300$\arcsec$, respectively. After standard data filtering, the total exposure time is 3.9~ks. The XRT spectrum was rebinned to have at a S/N of of least 5 per bin.

\begin{figure}
\centering
\includegraphics[width=0.49\textwidth]{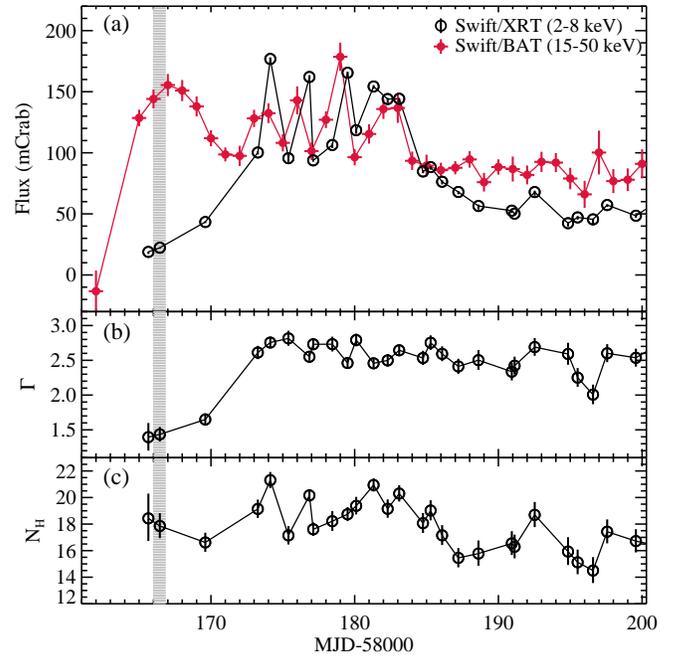}
\caption{(a) \swift\ monitoring of the outburst of \swiftsource. The BAT light curve is rescaled to the unit mCrab from count rates in the 15--50 keV band. The XRT light curve is converted to mCrab using calculated flux in the 2--8 keV band. (b) (c) Photon index $\Gamma$ and absorption column density $N_{\rm H}$ derived by fitting the XRT spectra with an absorbed power-law model. Gray shaded area marks the duration of the \nustar\ observation.
\label{fig:fig1}}
\end{figure}

\section{LIGHT CURVES}
\label{sec:sec3}
Evolution of the flux of \swiftsource\ since the beginning of the outburst observed by \swift/XRT and BAT is shown in Figure~\ref{fig:fig1}. The \swift/BAT light curve in daily averaged flux is from the \swift/BAT transient monitor \citep{swiftbat}. \swift/XRT data were reduced following the procedures described in Section~\ref{sec:sec2}, and the XRT flux was calculated by fitting the spectra with an absorbed power-law model, {\tt TBabs*powerlaw}. In this work, we perform all spectral fitting using XSPEC v12.9.0n \citep{xspec} with $\chi^2$ statistics, and adopt the cross-sections from \cite{crosssec} and abundances from \cite{wil00}. All parameter uncertainties are reported at the 90\% confidence level for one parameter of interest unless otherwise clarified. As shown in Figure~\ref{fig:fig1}b, the \nustar\ observation caught \swiftsource\ in the hard state before significant spectral softening occurred.

\begin{figure}
\centering
\includegraphics[width=0.49\textwidth]{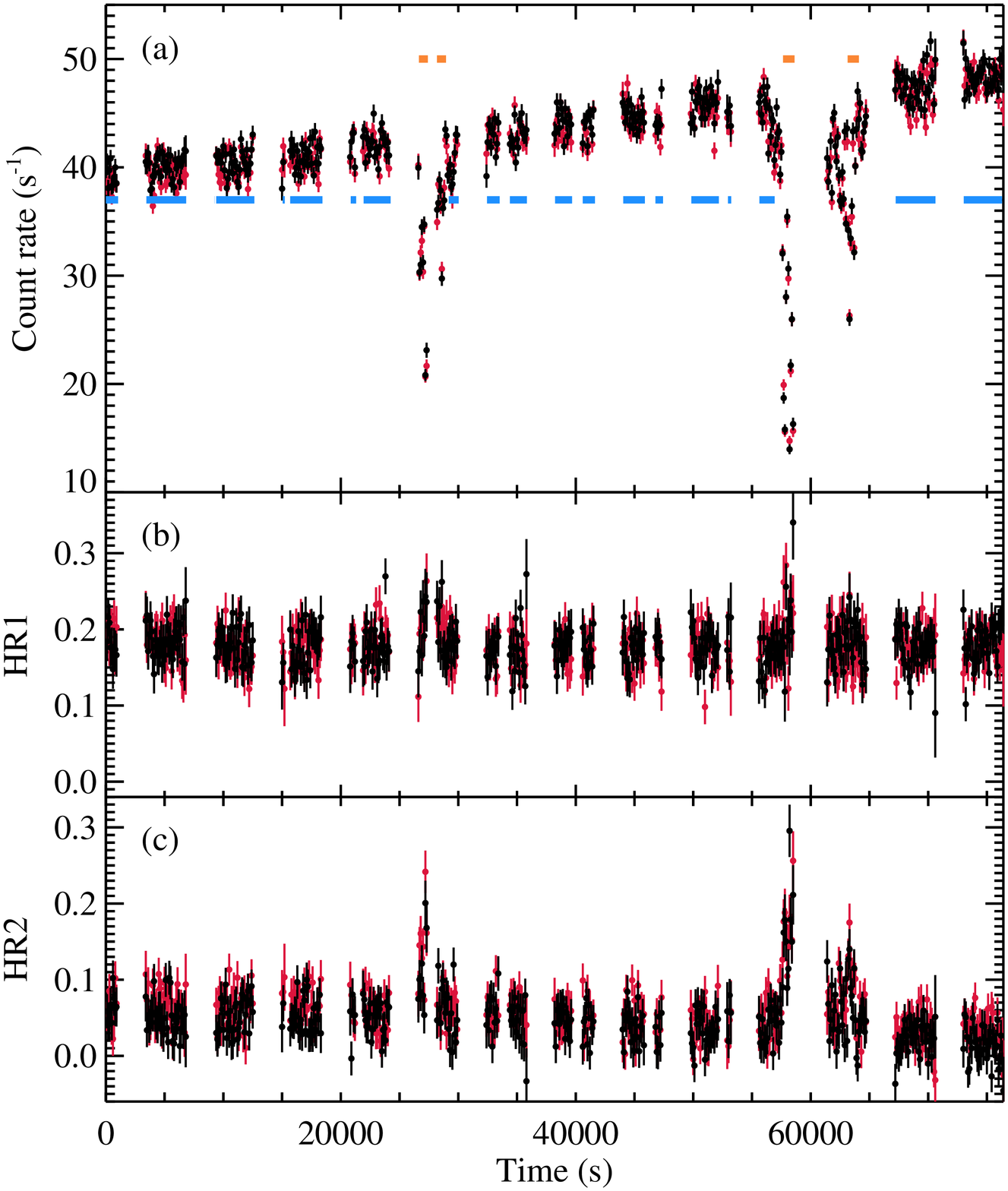}
\caption{(a) \nustar\ full band light curves of \swiftsource\ in 100s bins (FPMA in black, FPMB in red). Orange and blue lines mark the data intervals selected for the dip and persistent spectra, respectively. The orbital gaps in the light curves are due to occultations and SAA passages. (b) (c) Plots of corresponding hardness ratios. Hardness ratios HR1 and HR2 are defined as HR$1=(M-S)/(M+S)$ and HR$2=(H-M)/(H+M)$, where $S$, $M$, $H$ are the count rates in the energy bands of 3--6~keV, 6--10~keV and 10--79~keV, respectively.
\label{fig:fig2}}
\end{figure}

Three dips are evident in the \nustar\ light curve generated by the standard NUPRODUCTS procedure, where the \nustar\ full band count rate decreased by $\sim45\%-70\%$ at the dip minima (see Figure~\ref{fig:fig2}a). Dipping has been found to be periodic in the black hole binaries {GRO J1655--40} and {MAXI J1659--152}, and are associated with the binary orbital period in these systems \citep[][]{kuu98,kuu13}. It is possible that the dips observed in \swiftsource\ are also periodic, but cannot be confirmed with the limited exposure time of our \nustar\ observation. We extract source light curves from three energy intervals, 3--6~keV ($S$), 6--10~keV ($M$), 10--79~keV ($H$), and calculate hardness ratios defined as ${\rm HR1}=(M-S)/(M+S)$ and ${\rm HR2}=(H-M)/(H+M)$. The dips are clearly detected in all three energy bands. The hardness ratios are found to increase during the dips (see Figure~\ref{fig:fig2}b,c), indicating the dips are caused by increased absorption along the line of sight.

The source count rate was rising linearly during the \nustar\ observation. Despite the increase in count rate, the hardness ratio, HR1, basically remains constant outside of the dips. The overall change in HR2 is also relatively small (decreased by $\sim0.03$ by the end of the exposure), which in this case is mainly due to the change in the high energy cutoff. The cutoff energy, $E_{\rm cut}$, in black hole binaries has been observed to decrease monotonically with the increasing flux during the rising phases of their hard states \citep[e.g.,][]{joinet08,motta09}. We note here the corresponding changes in the absolute values of spectral parameters are subtle. Therefore, for spectral modeling, we only consider time-averaged spectra and separate the total exposure by dipping and non-dipping phases.

\section{SPECTRAL ANALYSIS}
\label{sec:sec4}
\subsection{Persistent spectra}
We first consider time-averaged spectra excluding the dips, which we refer to as the persistent spectra henceforth. Although the exposure of \swift/XRT is much shorter than that of the \nustar\ observation, \swift\ caught the second dip observed by \nustar, which we also excluded from the XRT data when extracting the persistent spectrum. The accumulated times for the persistent spectra are 30.9~ks, 31.4~ks and 3.6~ks for \nustar/FPMA, \nustar/FPMB and \swift/XRT, respectively. 

\begin{figure}
\centering
\includegraphics[width=0.49\textwidth]{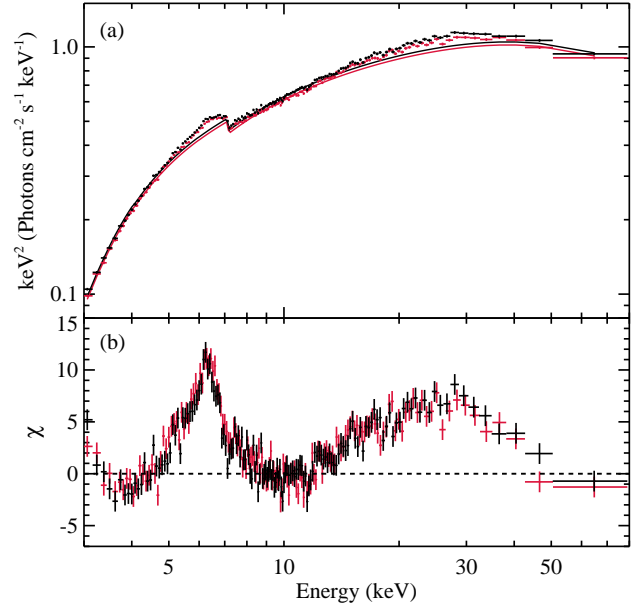}
\caption{(a) Persistent (non-dip) hard state spectra of \swiftsource\ observed by \nustar. FPMA and FPMB data are plotted in black and red, respectively. The energy spectra are folded with a cutoff power-law model. (b) Clear relativistic reflection features can be seen in the spectral residuals plotted in $\chi$. The spectra have been rebinned for display clarity.
\label{fig:fig3}}
\end{figure}

Relativistic reflection features are clearly detected in the \nustar\ spectra, including a broad Fe K$\alpha$ line and a Compton reflection hump peaking around 30~keV (see Figure~\ref{fig:fig3}). To highlight the disk reflection features, we first fit the \nustar\ spectra with an absorbed cutoff power-law model: {\tt TBnew*cutoffpl}, in XSPEC notation, only considering the energy intervals of 3--5, 8--12, and 40--79 keV. We note that in the residual plot, there is some indication for an absorption line around 7~keV (see Figure~\ref{fig:fig3}b), near the absorption edge. Throughout this work, we use the {\tt TBnew}\footnote{\label{footnote}http://pulsar.sternwarte.uni-erlangen.de/wilms/research/tbabs/} model to account for neutral absorption, and fix all abundances at solar in {\tt TBnew}. This updated version of the {\tt TBabs} model better characterises the shape of the absorption edges, which could be relevant here as the source has a high absorption column density. This approximate fit requires an absorption column density of $N_{\rm H}\sim1.3\times10^{23}$~cm$^{-2}$, a photon index of $\Gamma\sim1.3$, and a relatively low cutoff energy of $E_{\rm cut}\sim55$~keV. 

To obtain broad-band X-ray coverage so that spectral parameters can be better constrained, we jointly fit the \nustar\ and the \swift/XRT spectra. We allow the cross-normalization constants to vary freely for \nustar/FPMB and \swift/XRT, and fix the value at unity for \nustar/FPMA. We use XRT data above 1~keV, as low energy spectral residuals have been reported in heavily absorbed sources observed in the WT mode\footnote{\label{footnote}http://www.swift.ac.uk/analysis/xrt/digest\_cal.php}. Also \nustar\ data below 4~keV are ignored during the spectral fitting to avoid possible calibration uncertainties near the low energy end of the band pass. We use the model {\tt relxilllpCp} in the {\tt relxill} model family \citep{relxilla,relxillb} to physically model the relativistic reflection spectra. The {\tt relxilllpCp} model uses an idealized lamp-post geometry for the illuminating source commonly referred to as the corona. In this model, the corona is approximated as a point source located on the spin axis of the black hole above the accretion disk. The reflection fraction, $R_{\rm ref}$, can be self-consistently calculated in the lamp-post model based on the geometry assumed, which helps to reduce the parameter space. It uses the thermal Comptonization model {\tt nthcomp} \citep{nthcomp1,nthcomp2} as the input continuum, which provides a generally different spectral curvature from the phenomenological {\tt cutoffpl} model, especially in the case of a low high-energy cutoff. For all spectral fitting with {\tt relxilllpCp}, we fix the outer radius of the accretion disk $R_{\rm out}$ at 400 $r_{\rm g}$ ($r_{\rm g}\equiv {\rm GM}/c^2$ is the gravitational radius), and set the value of reflection fraction $R_{\rm ref}$ to be self-consistently determined by the model given the combination of the black hole spin parameter $a^*$, the inner radius of the accretion disk $R_{\rm in}$ and the lamp-post height $h$. The inner disk radius $R_{\rm in}$ and the black hole spin parameter $a^*$ are degenerate, as they both control the effective inner accretion disk radius. Therefore, for simplicity, during the spectral fitting we assume that the inner disk extends down to the ISCO by fixing $R_{\rm in}$ at the radius of the ISCO, and fit for the black hole spin as a free parameter (see Section~\ref{sec:sec6} for a discussion about disk truncation).

\begin{figure}
\centering
\includegraphics[width=0.50\textwidth]{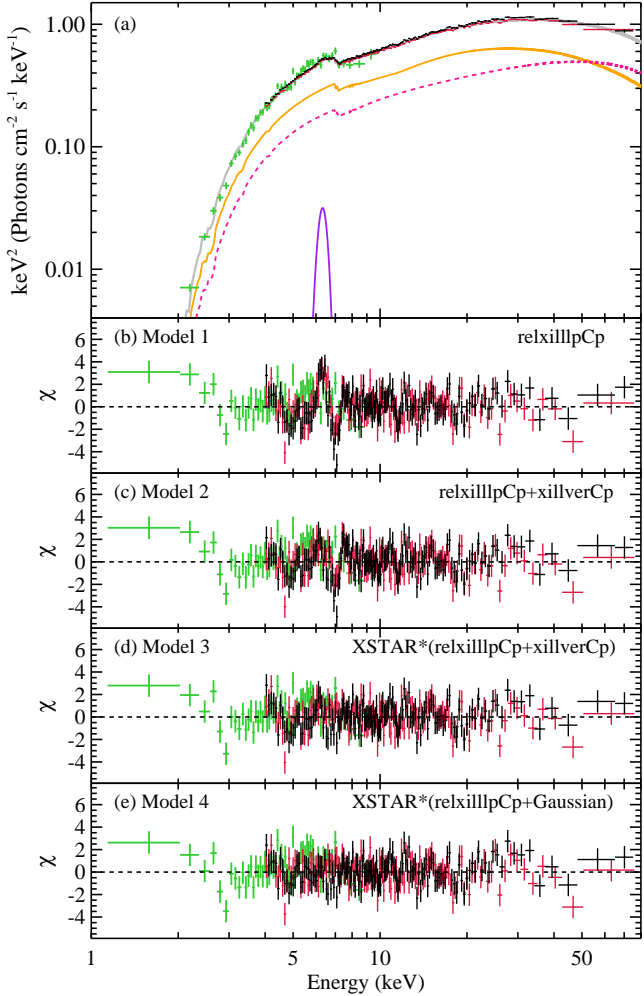}
\caption{(a) Persistent broad-band X-ray spectra of \swiftsource\ folded with the best-fit model (Model 4). The total model is marked in gray solid lines, with contributions from the Comptonization continuum (magenta), the disk reflection (orange) and the narrow Fe K line (purple). The \swift/XRT spectrum is plotted in green, and the same color scheme is used for the \nustar\ data as in previous plots. The spectra are rebinned for display clarity. (b)--(e) Spectral modeling residuals. \label{fig:fig4}}
\end{figure}

It is clear disk reflection alone {\tt TBabs*relxilllpCp} (Model 1) is not sufficient to describe the data (reduced chi-squared $\chi^2/\nu=1808.8/1599=1.13$, where $\nu$ is the number of degrees of freedom). Secondary features including a narrow iron line and possible absorption around 7~keV can be seen in the residual plot (see Figure~\ref{fig:fig4}b). A narrow iron line component on top of the broad Fe K profile has been found in a number of black hole binaries observed by \nustar\ \citep[e.g.,][]{walton16, walton17, xu_maxi18, tomsick18}, although the narrow line complex cannot be spectrally resolved. The origin of this spectral component is still controversial. Possible scenarios generally considered include distant reflection by a flared disk, line emission from the stellar wind of the donor star, or re-emission from an accretion disk wind if detected along with absorption features. We first tentatively fit for this feature using an unblurred reflection model {\tt xillverCp} \citep{xillver} to account for possible contribution from distant reflection, with the total model setup as {\tt TBabs*(relxilllpCp+xillverCp)} (Model 2). We assume the narrow iron line to be neutral by setting the ionization parameter in {\tt xillverCp} as log~$\xi=0$ ($\xi=4\pi F_{\rm x}/n$, where $F_{\rm x}$ is the ionizing flux, and $n$ is the gas density). We link other parameters in {\tt xillverCp} with the corresponding parameters in {\tt relxilllpCp}, only allowing the normalization of {\tt xillverCp} to vary freely. This results in an improvement of $\Delta\chi^2=93$ with one additional free parameter ($\chi^2/\nu=1715.8/1598=1.07$). Modeling the narrow iron line with {\tt relxilllpCp} instead results in a similar fit, and it yields a lower limit on the distance of the distant reflection component ($>300~r_{\rm g}$), which is far enough from the black hole to justify its low ionization state.

An apparent absorption line is still evident in the residuals of Model 2 (see Figure~\ref{fig:fig4}c). Fitting this feature with a simple Gaussian absorption line model brings an improvement of $\Delta\chi^2=67.5$ to the fit with three more parameters. It reveals that the line centroid lies at $E_{\rm abs}=7.03^{+0.04}_{-0.03}$ keV, with an equivalent width (EW) of $23^{+5}_{-4}$ eV. The absorption line is significantly detected at 7.5$\sigma$ confidence via a simple F-test. Strong absorption features from highly ionized iron are unusual for black hole binaries in the low/hard state, as they are considered as signatures of a powerful disk wind. Complications arise from the fact that the apparent absorption line is right at the Fe K edge. Assuming this is indeed an iron absorption line originating from a disk wind, if associated with blueshifted He-like Fe {\small XXV} (6.70~keV), it requires an extreme outflowing velocity of $0.049^{+0.006}_{-0.004}~c$ (14800$^{+1800}_{-1300}$ km~s$^{-1}$); alternatively, if this is identified with the more ionized H-like Fe {\small XXVI} (6.97~keV), it corresponds to a lower velocity of $0.009^{+0.006}_{-0.004}~c$ (2600$^{+1800}_{-1300}$ km~s$^{-1}$), but is still a high outflowing velocity for a disk wind launched by a stellar-mass black hole \citep[e.g.,][]{neil09,miller15_wind}. However, we note that in the latter case, the velocity shift could only be marginal when taking into account the absolute energy calibration uncertainty of \nustar\ \citep{madsen}.

\capstartfalse
\begin{deluxetable*}{ccccc}
\tablewidth{\textwidth}
\tablecolumns{5}
\tabletypesize{\scriptsize}
\tablecaption{Spectral Fitting of \swiftsource\ in the Hard State \label{tab:tab1}}
\tablehead{
\colhead{Component}       
& \colhead{Parameter}                           
& \colhead{Persistent (Model 3)}
& \colhead{Persistent (Model 4)}
& \colhead{Dipping (Model 5)}
} 
\startdata
{\textsc{tbnew}}           & $N_{\rm H}$ ($\rm \times10^{23}~cm^{-2}$)   &$1.85^{+0.07}_{-0.10}$      &$1.80\pm{0.08}$     &--               \\
\noalign{\smallskip}  
{\textsc{parcov*(tbnew*cabs)}}  & $N_{\rm H}$ ($\rm \times10^{24}~cm^{-2}$)   &\nodata                     &\nodata                    &$1.00^{+0.08}_{-0.07}$     \\
\noalign{\smallskip}
                           & $f_{\rm cov}~(\%)$                               &\nodata                &\nodata                    &$34\pm{1}$         \\
\noalign{\smallskip}                                                                                                                                        
{\textsc{xstar}}        & $N_{\rm H}$ ($\rm \times10^{22}~cm^{-2}$)    &$3.2^{+1.4}_{-1.0}$         &$2.0^{+1.2}_{-1.0}$        &--                \\
\noalign{\smallskip}                                                                                                              
                          & log~(${\xi}$) (log [$\rm erg~cm~s^{-1}$])                                  &$2.09^{+0.17}_{-0.07}$      &$2.00^{+0.27}_{-0.08}$                    &--                \\
\noalign{\smallskip}                                                                                                              
						  & $v_{\rm out}/c$                              &$0.077^{+0.006}_{-0.007}$           &$0.080^{+0.007}_{-0.014}$  &--                \\  
\noalign{\smallskip}
{\textsc{relxilllpCp}}    & $h$ ($r_{\rm g}$)                            &$<6.6$                      &$<3.8$                     &--                 \\
\noalign{\smallskip}                                                                                                              
				          & $a^{*}$ ($c{\rm J/GM}^{2}$)                      &$0.92^{+0.04}_{-0.06}$      &$>0.96$                    &--                  \\                                                                                                  
\noalign{\smallskip}
				          & $i$ ($^\circ$)                               &$61^{+3}_{-4}$              &$64^{+2}_{-3}$             &--                   \\
\noalign{\smallskip}                                                                                                              
				          & $\Gamma$                                     &$1.69\pm{0.02}$             &$1.63\pm{0.02}$            &--                  \\
\noalign{\smallskip}                                                                                                              
                          & log~(${\xi}$) (log [$\rm erg~cm~s^{-1}$])      &$3.49^{+0.10}_{-0.08}$      &$3.47^{+0.19}_{-0.09}$     &--                   \\
\noalign{\smallskip}                                                                                                              
				          & {$A_{\rm Fe}$ (solar)}                       &$0.60^{+0.07}_{-0.09}$      &$0.91^{+0.54}_{-0.08}$     &--                     \\
\noalign{\smallskip}                                                                                                              
				          & {$kT_{\rm e}$~(keV)}                         &$27\pm{3}$              &$22\pm{1}$                 &--                      \\
\noalign{\smallskip}                                                                                                              
                          & $R_{\rm ref}$                                &2.19                        &3.25                       &--                       \\
\noalign{\smallskip}                                                                                                              
                          & {Norm ($10^{-2}$)}                           &$1.5^{+1.2}_{-0.7}$         &$1.7^{+0.1}_{-0.4}$        &--            \\
\noalign{\smallskip}                                                                                                              
{\textsc{gaussian}}       & $E_{\rm emi}$~(keV)                               &\nodata                     &$6.30\pm{0.04}$     &--                        \\
\noalign{\smallskip}                                                                                                              
                          & $\sigma$~(keV)                               &\nodata                     &$0.23^{+0.08}_{-0.06}$     &--                        \\
\noalign{\smallskip}                                                                                                              
                          & Norm (10$^{-4}$)                             &\nodata                     &$5.7^{+1.4}_{-1.1}$        &--                        \\						  
\noalign{\smallskip}                                                                                                              
{\textsc{xillverCp}}      & Norm (10$^{-2}$)                             &$0.18\pm{0.05}$      &\nodata                    &--                        \\
\noalign{\smallskip}                                                                                                              
\hline                                                                                                                            
\noalign{\smallskip}                                                                                                              
                          & $\chi^2/{\nu}$                               &1650.0/1595                 &1625.7/1593                &307.6/287               
\enddata  
\tablecomments{
There is no error estimation for the reflection fraction parameter, $R_{\rm ref}$, as it is self-consistently calculated by the model in the lamp-post geometry. In model 5, we add one extra absorption component, {\tt parcov*(TBnew*cabs)}, to Model 4 in order to fit the dip spectra. Corresponding parameters in Model 5 are fixed at the best-fit values from Model 4.}                                                                                                           
\end{deluxetable*}

\begin{figure}
\centering
\includegraphics[width=0.49\textwidth]{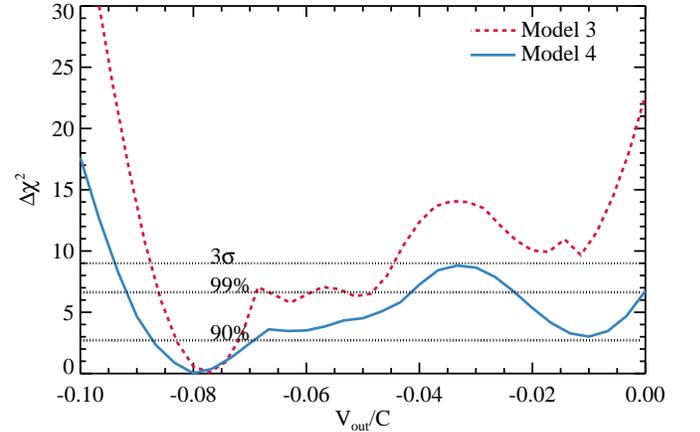}
\caption{$\Delta\chi^2$ plot of the velocity shift of the ionized absorber. The dashed lines mark the 90\%, 99\%, and 3$\sigma$ confidence levels for one parameter of interest. \label{fig:fig5}}
\end{figure}

In order to obtain a more physical interpretation, we fit for the absorption feature with an {\small XSTAR} table model \citep{xstar}: the full model is setup as {\tt TBabs*XSTAR*(relxilllpCp+xillverCp)} (Model 3). The {\small XSTAR} photoionization grid is customized for the hard state of \swiftsource\ using a power-law input spectrum with the photon index of $\Gamma=1.7$. We also assumed a gas density of $n=10^{14}$~cm$^{-3}$ following \cite{miller15_wind}, a turbulent velocity of 1000~km~s$^{-1}$ and a source luminosity of $L=10^{38}$~erg~s$^{-1}$. The grid covers the parameter space of $1.5\le{\rm log}(\xi)\le4.5$, $10^{21}~{\rm cm}^{-3}\le N_{\rm H}\le10^{23}~{\rm cm}^{-3}$ and $0.1\le A_{\rm Fe}\le10.0$, with the velocity shift as a free parameter. We link the iron abundance of the {\small XSTAR} model (in solar units), $A_{\rm Fe}$,  with that of the disk reflection component. The {\small XSTAR} model successfully accounts for the absorption feature with an ionization parameter of ${\rm log}(\xi)=2.09^{+0.17}_{-0.07}$, and a very high outflowing velocity shift of $v_{\rm out}=0.077^{+0.006}_{-0.007}~c~(22800\pm1800$~km~s$^{-1})$. We stress that obtaining an unique constraint on the ionization parameter and the velocity shift simultaneously is difficult here. Since only one absorption line can been clearly seen in the Fe K band, the ionization parameter and the velocity shift of the {\small XSTAR} grid are degenerate. At $3\sigma$ confidence level, a velocity of $v_{\rm out}$ at $\sim0.11~c$~(3300~km~s$^{-1}$) can still be considered as marginally acceptable (see Figure~\ref{fig:fig5} for the constraint on $v_{\rm out}$). It indicates that associating the absorption feature with more highly ionized Fe K species cannot be completely ruled out. The heavily absorbed nature of the source plus the short XRT exposure time precludes the possibility of detecting corresponding absorption lines in the soft end of the spectra. Model 3 provides an overall satisfying fit of the spectra, the results indicate that the source is a rapidly spinning black hole with the spin parameter of $a^{*}=0.92^{+0.04}_{-0.06}$, and is viewed at a high inclination angle of $i=61^{+3}_{-4}{^\circ}$.

In an attempt to further improve the fit, we replace the {\tt xillverCp} component with a simple {\tt Gaussian} emission line model (Model 4), since there is no clear indication that the narrow core of the Fe K$\alpha$ profile arises from reflection. This results in a better fit by $\Delta\chi^2=25$, with the difference mainly lying at the red wing of the narrow iron line around 6~keV (see Figure~\ref{fig:fig4}). The inclusion of an extra disk blackbody component ({\tt diskbb}) is not statistically required by the data. As the spectra display no obvious bump-like feature at low energies, most likely due to the high absorption column density, the strength or the temperature of the thermal disk emission cannot be well constrained. Although thermal emission from the inner accretion disk is physically required to explain a non-truncated disk, including a {\tt diskbb} model does not significantly affect the determination of other parameters during the spectral modeling. Therefore, we consider Model 4 as our best-fit model for the persistent broad-band X-ray spectra of \swiftsource.

Broadly similar results are obtained with Model 4: a low coronal height of $h<3.8~r_{\rm g}$ and a high spin of $a^{*}>0.96$, which lead to a high reflection fraction of $R_{\rm ref}=3.25$ calculated self-consistently by the model, signaling strong light-bending effects at the vicinity of the black hole. A high inclination angle of $i=64^{+2}_{-3}{^\circ}$ is measured for the inner accretion disk, which is sensitive to the blue wing of the broad iron line. Similar results are obtained if $R_{\rm ref}$ is left to vary freely, and an equally high black hole spin is measured, indicating that the high reflection fraction is indeed required by the data. We note that consistent results are obtained if we replace the lamp-post model used here ({\tt relxilllpCp}) with the model assuming a broken power-law coronal emissivity profile ({\tt relxillCp}); the best-fit values of the key physical parameters (black hole spin, disk ionization and inclination) are all consistent within errors. The constraint on the velocity shift of the {\small XSTAR} absorption component with Model 4 is weaker compared to that with Model 3. By allowing both line the width and centroid to vary in the {\tt Gaussian} emission line model, more flexibilities are brought to the spectral modeling near the Fe K edge. However, Model 4 still statistically prefers a considerably large outflowing velocity for the ionized absorber (see Figure~\ref{fig:fig5}). The most noticeable difference between Model 4 and 3 is the iron abundance $A_{\rm Fe}$ and the photon index $\Gamma$ (see Table~\ref{tab:tab1} for the list of spectral modeling parameters). $A_{\rm Fe}$ is driven by the relative flux in the Fe K band and of the Compton hump. Different from the distant reflection model, the narrow iron line component in Model 4 does not contribute to the flux in the high energy band ($>10$~keV). This would cause a small change in the overall spectral shape, resulting in corresponding changes in $\Gamma$ and $A_{\rm Fe}$. 

The fit confirms that the source is highly absorbed. The best-fit neutral absorption column density, $N_{\rm H}=1.81^{+0.07}_{-0.04}\times10^{23}$~cm$^{-2}$, is consistent with the measurement reported from initial analysis of the \swift/XRT data ($N_{\rm H}=(1.9\pm0.5)\times10^{23}$~cm$^{-2}$,  \cite{lien18}). This value is well in excess of the expected Galactic absorption in the direction of \swiftsource, $N_{\rm H, Gal}=1.55\times10^{22}$~cm$^{-2}$ \citep{kal05}. Therefore, the extra absorption is intrinsic to the source, and mostly likely originates from obscuring material near the orbital plane of the system. The parameters of the {\tt nthcomp} continuum are well constrained: a power-law index of $\Gamma=1.63\pm{0.01}$, indicating the source is in the hard state; and a relatively low coronal temperature of $kT_{\rm e}=22\pm1$~keV (the value here is as observed, not corrected for gravitational redshift at the vicinity of the black hole), which is determined by the exponential cutoff at the high energy end of the spectrum. Similar low coronal temperatures or equivalently low high-energy cutoffs have been previously measured by \nustar\ in black hole binaries during their bright hard states \citep[e.g.,][]{miller13,miller15,xu_maxi18}. 

\subsection{Dip spectra}
\begin{figure}
\centering
\includegraphics[width=0.49\textwidth]{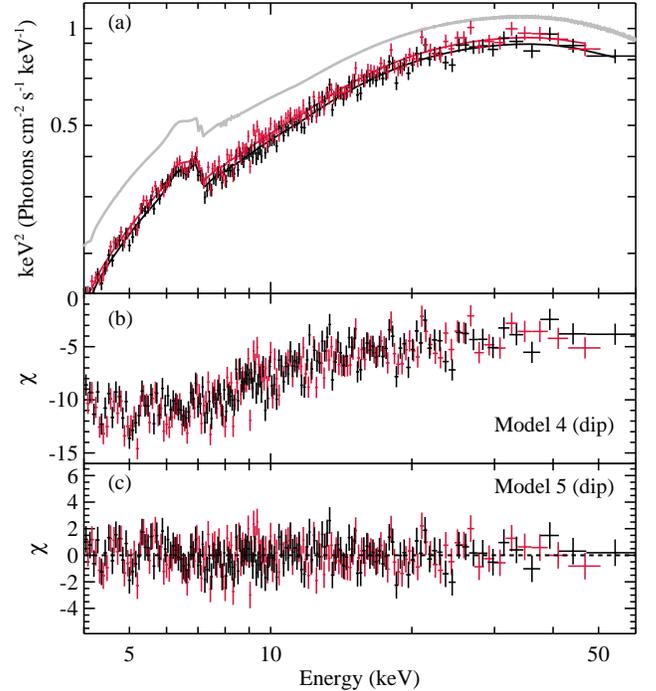}
\caption{(a) Residuals of the dipping spectra compared to Model 4 (best model for the persistent spectra). (b) Residuals of the dipping spectra from the best-fit model (Model 5). The spectra have been rebinned for display clarity.
\label{fig:fig6}}
\end{figure}

We extracted the spectra of the three dips collectively to obtain the highest S/N data possible for spectral fitting. The dipping intervals were selected as the periods when the source count rate clearly deviates from the linearly increasing trend (see Figure~\ref{fig:fig2} for details). This results in an exposure time of 3.0~ks and 3.3~ks for FPMA and FPMB, respectively. To highlight the spectral difference during the dips, we plot the residuals of the dipping spectra when compared to the best-fit model for the persistent spectra (Model 4) in Figure~\ref{fig:fig6}b. It is clear that dip spectra is harder, especially below 10~keV, suggesting that the spectral change is caused by absorption. 

An increase in the neutral absorption column density and changes in photon-ionized absorbers are usually invoked to explain spectral change during dipping intervals \citep[e.g.,][]{diaz06,shida13}. Evidence have also been found for progressive covering of the accretion disk corona as the dip progresses, during which the covering factor of the absorber was measured to increase,  approaching the maximum value at the dip minimum \citep[e.g.,][]{smale02, church04}. Our best-fit model for the dip spectra adds one extra partial covering neutral absorber to the persistent spectral model (Model 4). During the dips, electron scattering by the absorber could reduce the Comptonized emission from the corona, producing a nearly energy independent decrease in intensity  \citep[e.g.,][]{parmar86,smale02}. Therefore, we also include a {\tt cabs} model to account for the effects of absorbing material scattering X-ray photons away from the line of sight, and link the absorption column density in the  {\tt cabs} model with that of the partial covering absorber. The full model setup is {\tt TBnew*(partcov*}{\tt (TBnew*cabs))*}{\tt XSTAR*} {\tt (relxilllpCp+Gaussian)} (Model 5). Corresponding parameters in Model 5 are fixed at the best-fit values from Model 4, with only two parameters left to vary freely. The dip spectra can be adequately modeled after including the extra absorption component ($\chi^2/\nu=307.6/287=1.07$, with no obvious residuals in Figure~\ref{fig:fig6}c). The best fit requires a neutral absorber with the absorption column density of $N_{\rm H}=(1.00^{+0.08}_{-0.07})\times10^{24}$~cm$^2$ and a partial covering factor of $f_{\rm cov}=(34\pm{1})\%$. The fit confirms that the spectral changes during the dips can be fully accounted for by photoelectric absorption and Compton scattering from an additional neutral absorber.

\section{LOW-FREQUENCY QPO}
\label{sec:sec5}

\begin{figure*}
\centering
\includegraphics[width=0.95\textwidth]{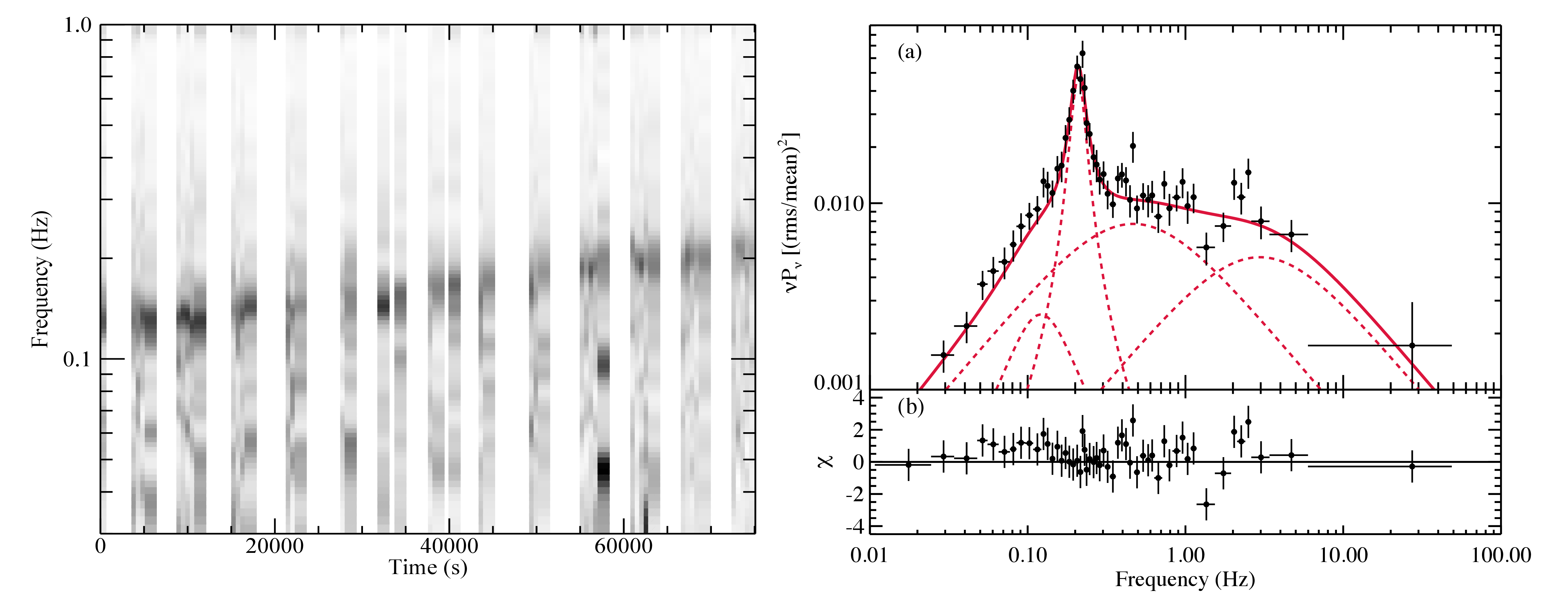}
\caption{Left: Dynamical power spectrum of \swiftsource\ from the \nustar\ observation. Low-frequency noise is present during the dipping period. Right: Sample power spectrum extracted from the last two orbits fitted with a multi-Lorentizan model. The power spectrum here have been rebinned for display clarity. 
\label{fig:fig7}}
\end{figure*}

For \nustar\ timing analysis, we first applied barycenter correction to the event files, transferring the photon arrival times to the barycenter of the solar system using JPL Planetary Ephemeris DE-200. The cross-power density spectra (CPDS) were generated using MaLTPyNT \citep{timingcode} following the standard procedure. CPDS measures the signals in phase between the two \nustar\ modules (FPMA and FPMB), which helps to reduce dead time distortions to the power density spectra \citep{nustartiming}. We selected a light curve binning of 2$^{-8}$~s and calculated the CPDS in 512~s intervals. The power spectra were generated using the root-mean-square (rms) normalization, and were geometrically rebinned by a factor of 1.03 to reach nearly equally spaced frequency bins in the logarithmic scale.

A low-frequency QPO is clearly detected by \nustar. The QPO frequency steadily increased from $\sim0.14$~Hz to $\sim0.21$~Hz during the \nustar\ exposure, as shown in the dynamical power spectrum (Figure~\ref{fig:fig7}, left panel). We extract an example CPDS from only the last two orbits, and fit it in XSPEC with a unity response file. We use a multi-Lorentzian model, which is commonly used to fit the power spectra of black hole binaries \citep[e.g.,][]{nowak00}. Our model for the power density spectrum includes four Lorentz functions: two Lorentzians with the centroid frequency fixed at zero to fit for the underlying continuum; one Lorentzian for the QPO, and one for the possible sub-harmonic with the frequency linked with half the fundamental QPO frequency. The sample power spectrum can be adequately fitted with no obvious structure left in the residuals, $\chi^2/\nu=292.6/243=1.20$ (see Figure~\ref{fig:fig7}, right panel). The QPO peak averaged over the last two orbits can be well constrained to be at the frequency of $0.207^{+0.003}_{-0.002}$~Hz, with a Q-value of $4.5\pm0.9$ and a fractional rms amplitude of $12.7\pm0.8\%$. The low QPO frequency, the increasing trend of the centroid frequency with the rising source flux, and the underlying noise continuum in the power density spectrum are consistent with type-C QPOs during hard states of black hole binary outbursts \citep[][]{wijn99,cas04,cas05}.

\section{DISCUSSION}
\label{sec:sec6}
In this work, we performed an X-ray spectral and timing analysis of the recently discovered X-ray transient \swiftsource, using data from joint \nustar\ and \swift\ observations. The shape of the broad-band X-ray spectral continuum and the detection of a low-frequency QPO is similar to the properties of black hole binary in the hard state. Given that no coherent pulsation or type I X-ray burst is detected, this suggests the source is a strong black hole binary candidate. From spectral modeling, the highly absorbed nature of \swiftsource\ is confirmed, with the neutral absorption column density well measured as $N_{\rm H}=(1.80\pm0.08)\times10^{23}$~cm~$^{-2}$. The hard photon index of $\Gamma=1.63\pm0.02$ and the simultaneous detection of a jet in the radio band \citep{russ18} suggests \swiftsource\ was in the hard state when the observations were taken. It was most likely in the bright phase of the hard state, as the source was reported to enter the hard-intermediate state four days later \citep{beri18}.  The averaged non-dip flux is about $30$~mCrab in the 2--10~keV band during the \nustar\ exposure, and the absorption corrected source luminosity is $L_{\rm x, 0.1-500~keV}=3.9\times10^{37}\times(D/8~{\rm kpc})^2$~erg~s$^{-1}$. Currently, the distance, $D$, to \swiftsource\ is unknown and the high absorption in the ISM along the line of sight makes it difficult to identify the optical counterpart.

Strong disk reflection features are shown in the \nustar\ spectra. Modeling the reflection features with the self-consistent disk reflection model {\tt relxilllpCp} indicates that the central black hole is rapidly spinning with the spin parameter of $a^*>0.96$, and the inner disk is viewed at a high inclination angle of $i=64^{+2}_{-3}{^\circ}$. The inclination angle measured is consistent with that fact that absorption dips are detected in the light curves. 

During spectral modeling of the reflection spectra in Section~\ref{sec:sec5}, we made the assumption that the inner accretion disk extends to the ISCO, which is not necessary true considering the possibility of a truncated disk in the hard state. The black hole spin and the inner disk radius are essentially degenerate parameters. The spin parameter is determined by measuring the location of the ISCO, which decreases monotonically from 9 $r_{\rm g}$ for an extremely retrograde spinning black hole to 1.235 $r_{\rm g}$ for a black hole with a maximum positive spin. Therefore, by assuming the disk extends to the ISCO rather than being truncated, we would be conservatively fitting for a lower limit of the black hole spin. Likewise, a high black hole spin obtained this way rules out the possibility of significant disk truncation. If we fix the spin parameter at the maximum value of $a^*=0.998$ and instead fit for $R_{\rm in}$, the best-fit value for the inner disk radius is $R_{\rm in}=1.7^{+0.2}_{-0.1}~r_{\rm g}$, with other parameters remaining at basically identical values. And solutions with $R_{\rm in}>3.3$~$r_{\rm g}$ can be ruled out at a 5$\sigma$ confidence level. 

A nearly maximal spin and the lack of significant disk truncation further support the hypothesis that \swiftsource\ contains a black hole primary. In the case of neutron stars, the Schwarzschild metric is normally a good approximation, and the neutron star spin parameter is expected to be less than 0.3 \citep[e.g.,][]{gal08}. Also in neutron stars the disk must be truncated either at the surface of the star or at its magnetospheric radius, with the typical measured truncation radius as $6-15~r_{\rm g}$ \citep[e.g.,][]{cack10, miller13, ludlam17}, which significantly exceeds the constraint on the inner disk radius we obtained for \swiftsource. The origin of QPOs in black hole binaries is still unclear despite being studied for decades. The lack of disk truncation plus the simultaneous detection of a type-C QPO in the hard state of a black hole binary candidate by \nustar\ is in tension with the Lense-Thirring precession
 model for QPOs, one of the currently promising physical models to explain the origin of low-frequency QPOs, as in the Lense-Thirring precession
 model the inner accretion disk is generally required to be truncated \citep[][]{stellar98,stellar99,ingram09}.

An apparent absorption line is detected in the Fe K band centered at $7.03^{+0.04}_{-0.03}$~keV. If associated with ionized iron absorption, this implies the unusual presence of a disk wind in the hard state of a black hole binary. As only one absorption feature is clearly detected, the velocity shift and the ionization state of the absorber cannot be unambiguously determined, similar to the case of 4U 1630--472 discussed in \cite{king14}. The best fit favors a lower ionization state and a very high outflowing velocity, comparable to the extreme wind velocity reported in the black hole binary candidate IGR J17091--3624 \citep{king12}. In the case of an extreme disk wind, the relatively narrow Fe K emission line component required in the spectral modeling could be a part of a P-Cygni profile, arising from redshifted ionized iron line emission in the wind. However, we stress that an alternative solution of a higher ionization and moderate outflowing velocity cannot be rule out from the spectral analysis, where the velocity shift of the ionized absorber is actually close to the absolute energy calibration uncertainty of \nustar. \swiftsource\ is an interesting source for studying the equatorial disk winds in binary systems. Future observations with high spectral resolution would be helpful to resolve this uncertainty.

\acknowledgments{
We thank the referee for helpful comments that improved this work. D.J.W. acknowledges support from STFC Ernest Rutherford Fellowship. This work was supported under NASA contract No.~NNG08FD60C and made use of data from the \nustar\ mission, a project led by the California Institute of Technology, managed by the Jet Propulsion Laboratory, and funded by the National Aeronautics and Space Administration. We thank the \nustar\ Operations, Software, and Calibration teams for support with the execution and analysis of these observations. This research has made use of the \nustar\ Data Analysis Software (NuSTARDAS), jointly developed by the ASI Science Data Center (ASDC, Italy) and the California Institute of Technology (USA).}

\bibliographystyle{yahapj}
\bibliography{j1658.bib}
\end{document}